\documentclass[12pt]{article}
\usepackage[english]{babel}
\usepackage{graphicx}
\usepackage{framed}
\usepackage[normalem]{ulem}
\usepackage{amsmath}
\usepackage[ruled]{algorithm2e}
\usepackage{amsthm}
\usepackage{amssymb}
\usepackage{amsfonts}
\usepackage{hyperref}
\hypersetup{
colorlinks=true,
linkcolor=blue,
citecolor=blue
}
\usepackage[noblocks]{authblk}
\usepackage{bm}
\usepackage{booktabs}
\usepackage{threeparttable}
\usepackage{multicol}
\usepackage{indentfirst}
\usepackage{cite}
\usepackage{enumerate}
\usepackage{color}
\usepackage[utf8]{inputenc}
\usepackage[top=1 in,bottom=1in, left=1 in, right=1 in]{geometry}
\usepackage{underscore}

\theoremstyle{definition}

\date{2019.08.25}
\begin{document}
\title{{\huge\textbf{Deep Learning for Stock Selection Based on High Frequency Price-Volume Data}}}

\author[$^{\ddag}$]{{Junming Yang$^{*}$}}
\author[$^{\dag}$]{Yaoqi Li$^{*}$}
\author[$^{\ddag}$]{{Xuanyu Chen$^{*}$}}
\author[$^{\ddag}$]{{Jiahang Cao$^{*}$}}
\author[$^{\ddag}$]{{Kangkang Jiang$^{*}$}}
\affil[*]{Likelihood Technology}
\affil[$^{\dag}$]{University of California, Irvine \authorcr yaoqil1@uci.edu      deniselll877@gmail.com}
\affil[$^{\ddag}$]{Sun Yat-sen University \authorcr \{chenxy628, yangjm25, caojh7, jiangkk3\}@mail2.sysu.edu.cn}
\maketitle
\tableofcontents

\clearpage
\section{INTRODUCTION}
Training a practical and effective model for stock selection has been a greatly concerned problem in the field of artificial intelligence. Because of the uncertainty and sensitivity of the finance market, there are many factors which may influence the stock price such as significant events, society’s economic condition or political turmoils. Many scholars have been applied various methods of machine learning to find a fitting model for the stock price time-series data with nonlinearities, discontinuities, and high-frequency multi-polynomial components.\cite{hadavandi2010integration} To handle these complicated components and make a precise prediction, a lot of scholars choose to use machine learning to create a model.

\subsection{Related Work}
Shunrong Shen et al. proposed data from different global financial markets with SVM (support vector machine) and reinforce learning to predict stock index movements in the U.S. market. \cite{shen2012stock} Kai Chen et al. introduce the application of LSTM (Long Short-Term Memory) in stock index prediction by using low-frequency data. \cite{chen2015lstm} Hailin Li et al. compared actor-only and actor-critic reinforce learning methods for presenting reinforcement-oriented schemes to forecast short-term stock price movements. \cite{li2007short} Furthermore, Yue Deng et al. constructed a new model DDRL(deep direct reinforce learning) which combined direct reinforce learning, deep recurrent neural network, and fuzzy learning. \cite{deng2016deep}

\subsection{Background}
Even though some of the models from previous works have been achieved good performance in the U.S. market by using low-frequency data and features, training a suitable model with high-frequency stock data is still a problem worth exploring. There is another challenge for stock trading novices without experience in this field. Although there are existing low-frequency features created by some experts, constructing useful high-frequency features with high-level information is difficult for us. Moreover, many existing features which are calculated by the U.S. stock market index different from the China stock market. Therefore, we prefer to apply methods without constructing features by ourselves. In this paper, we introduce two machine learning algorithms LSTM (long short-term memory) and CNN (convolutional neural network) to find the most beneficial strategy of stock trading in China stock market.

\subsection{Summary}
Through the high-frequency price data of the past period (one day or several days), we construct two models which can predict the expected return rate of the stock on the day, and select the stock with the highest expected yield at the opening to maximize the total return. The organization of the rest paper is as follows: Section 2 presents general concepts, advantages and detailed constructions for the two models and our preparation of processing data. Section 3 describes hyperparameter selection, regularization and how we train our models. Section 4 compares the accuracy, thermodynamic diagram and the result of backtesting to analyze their advantages and weakness. At last, we summarize our work and indicate several future attempts.

\section{METHODOLOGY}
\subsection{Long Short-Term Memory}
One of a special kind of RNN(Recurrent Neural Network) is called Long Short-Term Memor(LSTM) which was first proposed by Hochreiter and Schmidhuber in 1997.\cite{hochreiter1997long} Gers et al. advanced its structure by introducing forget gate in 2000.\cite{gers1999learning} after a few years, H.Sak et al. and W.Zaremba et al. improved the framework of LSTM and added more details.\cite{sak2014long,zaremba2014recurrent} Nowadays, LSTM is widely used in the field of natural language processing and emotion analysis.

\subsubsection{Basic Theorem}
The traditional neural network always widely used in image processing and other problems without significant effects caused by time series. However, there are some situation depending on time series such as natural language processing and predicting stock price. In this process, we prefer to receive new information and also maintain preceding information because understanding earlier events can help us study current material. RNN address this issue by using networks with loops to allow information to preserve.

    \begin{figure}[h]

    \centering

    \includegraphics[width=.8\textwidth]{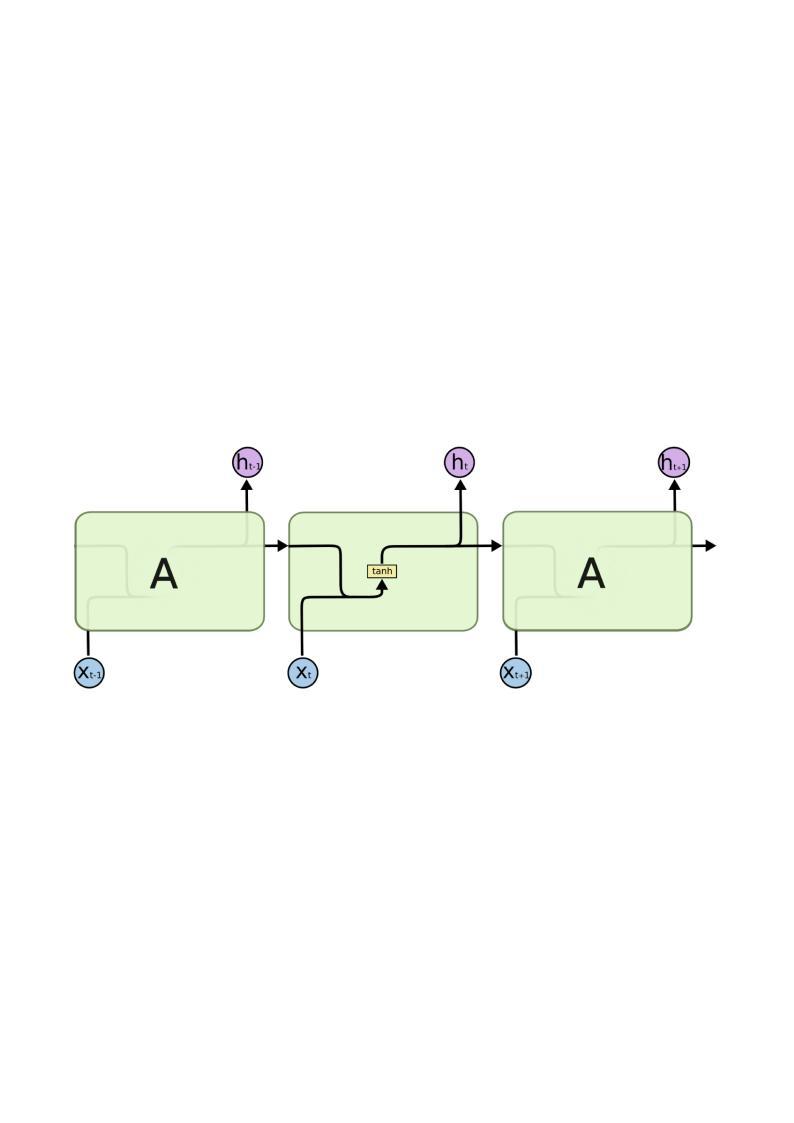}

    \caption{An unfolded Recurrent Neural Network}

    \label{rnn}

    \end{figure}

If we unfold RNN, we can get Figure \ref{rnn} a chain-like sequence of neural networks which are intimately related to time series.
Even though we can connect earlier information to the present task, we may need long-term memory to inform the understanding of current work. Considering two stocks with similar price tendency in the present week, according to data from this week, we are hesitated to buy which one. If we know in the last month, one of them hit limit up while another fell staying, then we can decide to buy the go up staying stock based on the understanding of information from previous days. Therefore, LSTM solves this problem and works tremendously well on time-series with long-term information problems.\cite{colahblog} According to the basic model of RNN, if we unfold the structure of one layer RNN, we can easily assume it as a deep feedforward neural network with one dense layer which shares the same weight. By appending more neural network layers and gates, LSTM adequate to handling long-term dependencies.

\subsubsection{Advantages of LSTM}
Comparing to other artificial neural networks, RNN has an advantage in figuring out problems which are sensitive to time series such as language translation and emotion analysis. Even though their purpose is to learn long-term dependency, some shreds of evidence prove that recurrent neural network troubled in learning and keeping long-term memory. Fortunately, by imitating the human being’s memory mechanism, LSTM solved this problem appropriately. To learn how to classify meaningful information and remove useless memory, LSTM separate unit state into long-term memory state and short-term memory state and add the gate system. Based on the previous study, LSTM is more effective than traditional RNN because not only can it converge quickly, also it can solve the problem of vanishing gradient and exploding gradient. Moreover, considering the feature that LSTM can maintain long-term memory, it is sensitive to the long-term dependency of data.

\subsubsection{Structure of LSTM}
    \begin{figure}[h]
        \centering
        \includegraphics[width=.8\textwidth]{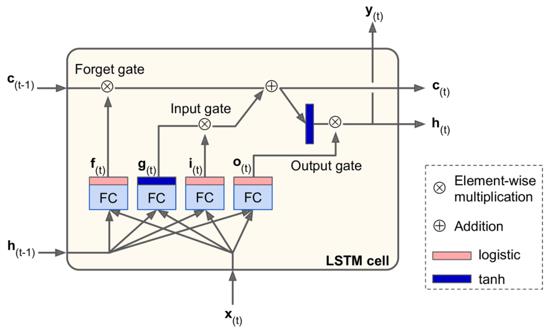}
        \caption{Basic Cell of Long Short Term Memory}
        \label{lstmintro}
    \end{figure}
In Figure \ref{lstmintro}, it shows that the LSTM unit state has two vectors: $\mathbf{h}_{(t)}$ and $\mathbf{c}_{(t)}$ which represent short-term memory state and long-term memory state. Each cell has three gates (fully connected layer) at $t$ time: $\mathbf{f}_{(t)}$ (forget gate), $\mathbf{i}_{(t)}$ (input gate) and $\mathbf{o}_{(t)}$ (output gate). The active function of the gates is the Logistic function whose output is in the range of $0$ to $1$. The gates uses element-by-element multiple to control how many data should be maintained.
\begin{quote}
\begin{itemize}
\item[$\bullet$]$\mathbf{h}_{(t)}$: short-term memory states
\item[$\bullet$]$\mathbf{c}_{(t)}$: long-term memory states
\item[$\bullet$]$\mathbf{x}_{(t)}$: a vector corresponding to input time series
\item[$\bullet$]$\mathbf{f}_{(t)}$: decide which part of $\mathbf{c}_{(t-1)}$ from the previous $t-1$ time can be stored in current $\mathbf{c}_{(t)}$
\item[$\bullet$]$\mathbf{g}_{(t)}$: analyze $\mathbf{x}_{(t)}$ and $\mathbf{h}_{(t)}$
\item[$\bullet$]$\mathbf{i}_{(t)}$: decide which part of the output returned by $\mathbf{g}_{(t)}$ should be added to current $\mathbf{h}_{(t)}$
\item[$\bullet$]$\mathbf{o}_{(t)}$: decide which part of current $\mathbf{c}_{(t)}$ should be regarded as current $\mathbf{h}_{(t)}$ and the output $\mathbf{y}_{(t)}$ of the cell at $t$ time
\end{itemize}
\end{quote}

Their computational formulas are as follows:
\begin{displaymath}
\begin{aligned}
\mathbf {i}_{(t)}&={\sigma}(
\mathbf {W}_{xi}^{T}\cdot \mathbf {x}_{(t)} +
\mathbf {W}_{hi}^{T}\cdot \mathbf {h}_{(t-1)} + \mathbf {b}_{i})\\
\mathbf {f}_{(t)}&={\sigma}(
\mathbf {W}_{xf}^{T}\cdot \mathbf {x}_{(t)} +
\mathbf {W}_{hf}^{T}\cdot \mathbf {h}_{(t-1)} + \mathbf {b}_{f})\\
\mathbf {o}_{(t)}&={\sigma}(
\mathbf {W}_{xo}^{T}\cdot \mathbf {x}_{(t)} +
\mathbf {W}_{ho}^{T}\cdot \mathbf {h}_{(t-1)} + \mathbf {b}_{o})\\
\mathbf {g}_{(t)}&={\tanh}(
\mathbf {W}_{xg}^{T}\cdot \mathbf {x}_{(t)} +
\mathbf {W}_{hg}^{T}\cdot \mathbf {h}_{(t-1)} + \mathbf {b}_{g})\\
\mathbf {c}_{(t)}&=\mathbf {f}_{(t)} \otimes \mathbf {c}_{(t-1)}
+ \mathbf {i}_{(t)} \otimes \mathbf {g}_{(t)}\\
\mathbf {y}_{(t)}&=\mathbf {h}_{(t)} = \mathbf {o}_{(t)}
\otimes \tanh(\mathbf {c}_{(t)})
\end{aligned}
\end{displaymath}

$\sigma$ represents the function as follows:
\begin{displaymath}
\sigma(x)=\frac{1}{1+e^{-x}}
\end{displaymath}

$\tanh$ represents the function as follows:
\begin{displaymath}
\tanh(x)=\frac{\sinh x}{\cosh x}=\frac{e^{x}-e^{-x}}{e^{x}+e^{-x}}
\end{displaymath}

The weight matrix $\mathbf {W}$ and the bias $\mathbf {b}$ will be generated by Back propagation algorithm.

\subsubsection{LSTM Model Construction}
\begin{figure}[htbp]
        \centering
        \includegraphics[width=.9\textwidth]{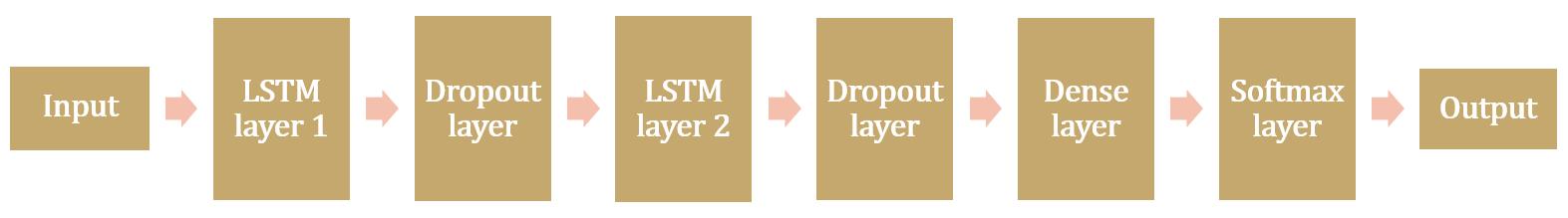}
        \caption{Long Short Term Memory Model Structure}
        \label{lstm}
    \end{figure}

Here, we use $\bm{x}_{it}$ to represent the feature vector of stock $i$ at time $t$. We treat a time series $\{\bm{x}_{it}\}\ (t=1,2,\ldots,T)$ as an instance to be the input of our LSTM network which can predict the profit $y$ of the next day. To consider this problem reasonably, we make this problem to be a classification problem. Hence, we create a classification model depended on LSTM network in Figure \ref{lstm}.
The LSTM layer of this model has been introduced carefully in the previous section. In the dropout layer, each cell has a certain probability to stop working in each training process to avoid over-fitting of the model. Therefore, when predicting, all neurons participate in the work to maximize the role of the model. Earlier experiments have confirmed that the addition of this layer has a good performance on regularization.\cite{hochreiter1997long} The Dense layer has no activation function and outputs the score of the probability that the predicted instance belongs to each category. The score is then converted to the corresponding probability by the softmax layer. In this process, softmax layer use the equation as follows:
\begin{displaymath}
P_{i}=\frac{e^{i}}{\sum_{j}e^{j}}
\end{displaymath}
In this equation, $P_{i}$ and $i$ represent the probability and score respectedly that the predicted instance belongs to certain category. The category with the highest probability is regarded as the prediction made by the model on the instance category.

\subsection{Convolutional Neural Network}
\subsubsection{Basic Theorem}
The convolutional neural network is an advanced neural network with the ability of multilayer back-propagation supervised learning networks which first proposed by Y. LeCun in 1995.\cite{lecun1995convolutional} CNN achieves some degree of distortion and deformation by using three architectural ideas: local receptive fields, shared weights, and spatial subsampling(convolution). Local receptive fields are inspired by the cortical structure of animal vision that only part of their neurons works in the process of sensing the external environment. The reason for using local receptive fields is that the correlation of the adjacent pixels in the image is relatively close and each cell only needs to perceive the local area without connecting to the whole picture. \cite{lawrence1997face} By applying the connecting method of shared weights and local receptive fields, CNN reduces the number of parameters.

Similar to other neural networks, CNN also has an input layer, an output layer, and several hidden layers. These three types of layers are the core components for implementing the feature extraction of CNN. By using the gradient descent method to adjust the weight parameters in the network layer by layer, the model minimizes the loss function and improves the accuracy of the network through frequent iterative training. The low hidden layer of the convolutional neural network is composed of a convolutional layer and a maximum pool sampling layer. And there are hidden layer and logistic regression classifier of the full-connected layer corresponding to the traditional multi-layer perceptron. The input of the first fully connected layer is always a featured image with spatial and temporal dependencies obtained by feature extraction from the convolutional layer and the sub-sampling layer. The last layer of the output layer is a classifier that can be classified using logistic regression or SVM(support vector machine) or other methods.

\subsubsection{CNN Model Construction}
As introduced above, through convolution between a convolution kernel and various regions of a two-dimensional input (e.g. a picture), Convolutional Neural Networks (CNN) can convert primary features of each region into higher-level features, serving as a feature extraction tool. Such function of CNN can be utilized to deal with our task. That is because the data input of a stock in one day is a two-dimensional one, one dimension being the set of features and the other being the time periods in a day; the input data are described in Table \ref{table2} from later section. When it comes to the shape of convolution kernels, it is for certain that we do not need to use the common shapes in image identification like 3$\times$3 or 5$\times$5. When dealing with high frequency price-volume data, the shape of convolution kernels should be designed according to specific purposes.

E. Hoseinzade and S. Haratizadeh introduced a CNN structure to make predictions of stock price with a set of low frequency features. CNN framework proposed by this paper is inspired by their structure. That is because our task and data structure are very similar to those of theirs. The difference is that they use low frequency data with time period being day instead of minutes and features being low frequency ones like EMA10. The framework of theirs will be demonstrated in details when comparing with ours in following sections.

\subsubsection{CNN Framework Proposed}
\paragraph{Convolutional Layer}
In the framework by E. Hoseinzade and S. Haratizadeh \cite{hoseinzade2019cnnpred},  the first layer is a convolution layer in which convolution kernels are utilized to extract high-level features. And in CNN framework proposed by this paper, the same first layer is used. Specifically, since we use 11 price-volume features under the frequency of 15 minutes, convolution kernels with shape of 1$\times$11 are utilized. Each of those filters covers all the daily variables and can combine them into a single higher-level feature (i.e. high-level features). Such kernels can construct different combinations of primary variables using this layer. It is also possible for the network to drop useless features by setting their corresponding weights in the kernels equal to zero. So, this layer works as a feature extraction or feature selection module.

After the search of hyperparameters, we eventually decide to use 40 kernels, which takes both the training cost of the model and the prediction effectiveness into account. Through this convolution layer, we obtain 40 high-level features, which are extracted from 11 primary features. And the two-dimensional data structure changes from 16$\times$11 (time-period $\times$ original features) to 16$\times$40 (time-period $\times$ high-level features).

\paragraph{Fully Connected Layers}
    \begin{figure}[htbp]
        \centering
        \includegraphics[width=.9\textwidth]{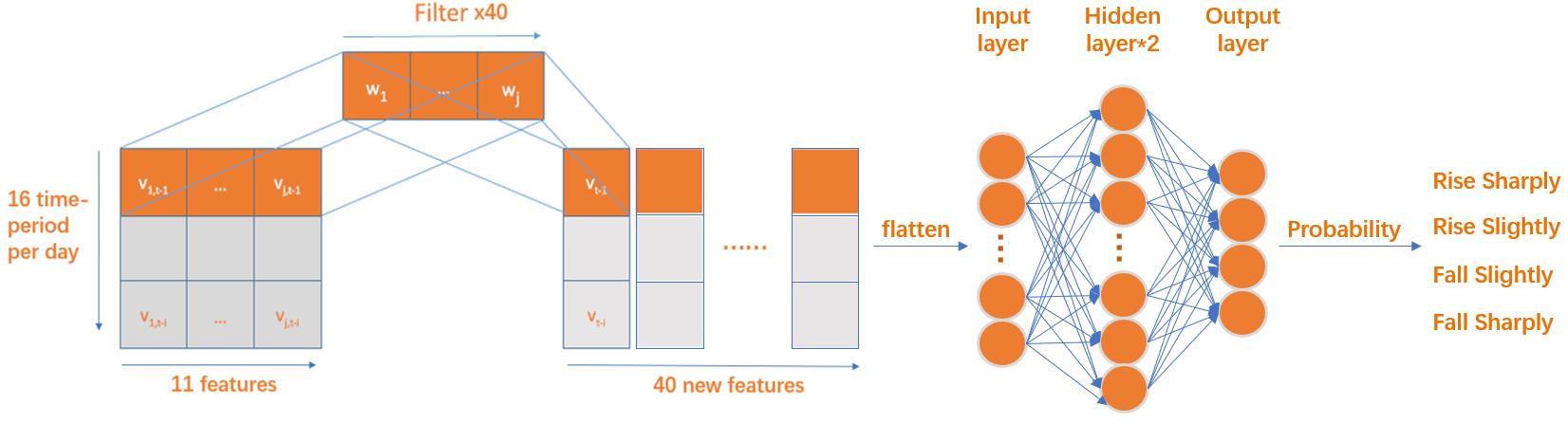}
        \caption{Convolutional Neural Network + 2 Dense Framework}
        \label{cnnframe}
    \end{figure}

In these layers, the 16$\times$40 data generated by the CNN layer are flattened into a final feature vector. This feature vector is then converted to a final prediction through 2 fully connected layers (2 hidden layers). And the output layer has 4 neurons, with SoftMax function utilized as the activation function, intended to give the probability of a big rise, a small rise, a big drop, a small drop of a stock respectively, and then calculate the expectation of its daily return today to make stock purchase advice. The whole framework is demonstrated in Figure \ref{cnnframe}

\section{EXPERIMENT}
\subsection{Data Processing}
\subsubsection{Sources of Data}
We have chosen the closing price, opening price, highest price, lowest price, trading volume, transaction amount, number of transactions, commission ratio, volume ratio, commission purchase, commission sale of the China A-share market. These 11 volume-price features are used as elements to describe the state of the stock. In order to train with market-represented stocks and reduce data inconsistency (such as stock suspension) and noise, we selected the constituents of the CSI300 Index, denoted as $I$, as the source of the sample data set. The model uses two types of data, every 15 minutes of data and every 120 minutes of data. The sample data is exhibited as in Table \ref{table2} and \ref{table3}.

\subsubsection{Time Period of Data}

\clearpage
\begin{figure}[htbp]
        \centering
        \includegraphics[scale=0.2]{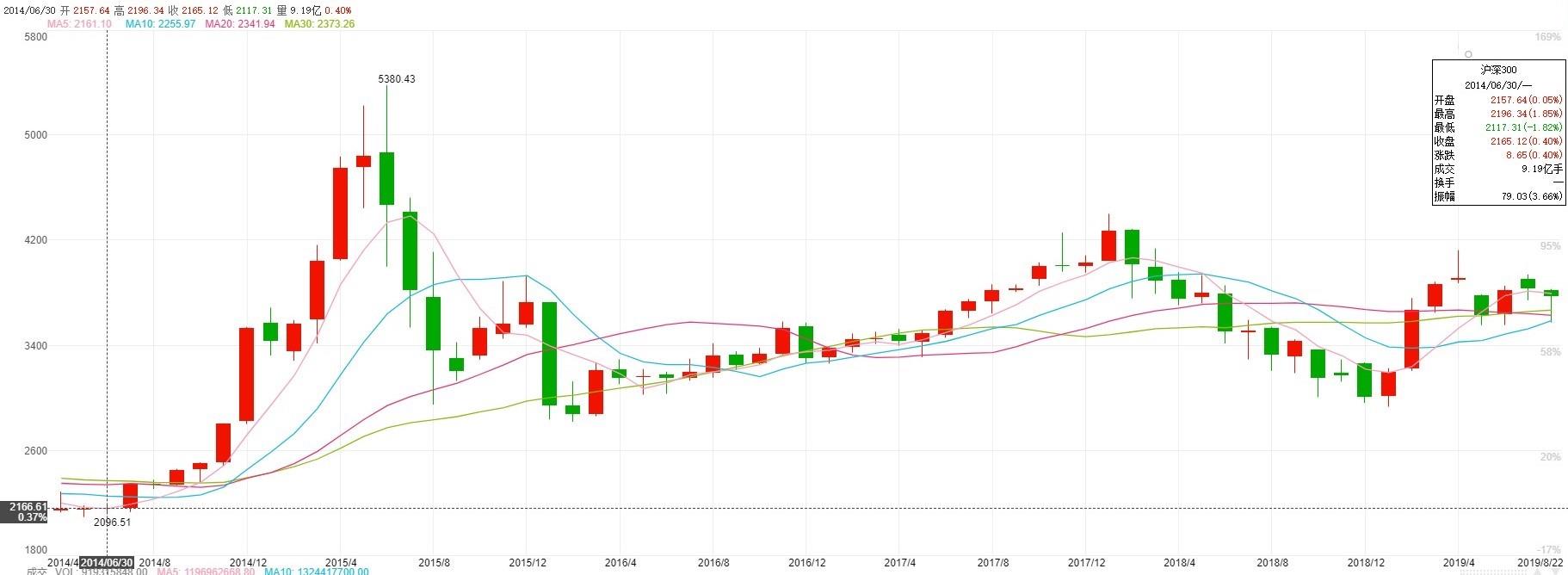}
        \caption{Trends of the CSI300 Index from July, 2014 to December, 2018}
        \label{trend}
    \end{figure}
As for sample set, we choose the data in Figure \ref{trend} from July 1,2014 to December 31, 2018, denoted as $M$, during which Chinese stock market witnessed periods of sharp rise, sharp fall, slight rise and slight fall, providing sufficient samples for each of our four labels.
Since our ability of computation is insufficient, when we adjusted the model, we only used data from January,2019 to May,2019, denoted as $N$.

\subsubsection{Data Normalization}
To speed up the convergence of the neural network and to eliminate the negative influence of the dimension of the feature data on the model [2], the feature vector $\bm{x}$ of each stock $i$ at each time $t$ is normalized by the equation as follows:
\begin{displaymath}
\begin{aligned}
\bm{x}&=(x_{1},x_{2},\ldots,x_{11})\\
\quad x'_{k}&=\left\{\begin{array}{ll}

\displaystyle \frac{x_{k}-\mathop{\min}\limits_{t\in T}\{x_{k}\}}
{\mathop{\max}\limits_{t\in T}\{x_{k}\}-\mathop{\min}\limits_{t\in T}\{x_{k}\}}

&\qquad\quad\textrm{if }\
\mathop{\max}\limits_{t\in T}\{x_{k}\}-\mathop{\min}\limits_{t\in T}\{x_{k}\}\neq 0\vspace{2ex}\\

0&\qquad\quad\textrm{if }\
\mathop{\max}\limits_{t\in T}\{x_{k}\}-\mathop{\min}\limits_{t\in T}\{x_{k}\}=0

\end{array}\right.\\
\bm {x}'&=(x'_{1},x'_{2},\ldots,x'_{11})\\
\forall i&\in I,\ t\in T,\ k=1,2,\ldots,11\\
\end{aligned}
\end{displaymath}
then convert $y$ from the rate of return to the category.

\subsubsection{Label Selection}
We divide the daily rate of return into four categories: a big rise, a small rise, a big drop, a small drop. To avoid the problem of category imbalance, we use the sample to estimate the whole dataset. By taking the daily rate of return of all A-shares from June, 2014, to December, 2018 (closing price minus opening price) as a sample $J$, we calculate its quartile as the division point of the classification.

\begin{table}[htbp]
	\footnotesize
	\renewcommand\arraystretch{1}
	\centering
	\caption{Labels selected}
	\label{table1}
	\begin{tabular}{lcccc}
		\toprule
		$\mathbf{Labels}$&$\mathbf{Fall\ Sharply }$&$\mathbf{Fall\ Slightly}$&$\mathbf{Rise\ Slightly}$&$\mathbf{Rise\ Sharply}$\\
		\midrule
		$\mathbf{Yields\ Interval}$&$-10\%\sim-1.06\%$&$-1.06\%\sim+0.04\%$&$+0.04\%\sim+1.32\%$&$+1.32\%\sim+10\%$\\
		\bottomrule
	\end{tabular}
\end{table}

\subsubsection{Division of Training Set and Test Set}
\begin{table}[ht]
	\scriptsize
	\renewcommand\arraystretch{1.2}
	\centering
	\caption{\footnotesize Data Example 1\ \ \ \ \ \ \ \ (Stock: SH600000\ \ \ \ Date: Jan $2^{th}$, 2019)}
	\label{table2}
	\begin{threeparttable}
		\begin{tabular}{cccccccccccc}
			\toprule
			$\mathbf{Time}$&$\mathbf{Close}$&$\mathbf{Open}$&$\mathbf{Highest}$&$\mathbf{Lowest}$&$\mathbf{Volume}$&$\mathbf{Amount}$&$\mathbf{NOT}$\tnote{1}&$\mathbf{CR}$\tnote{2}&$\mathbf{VR}$\tnote{3}&$\mathbf{CP}$\tnote{4}&$\mathbf{CS}$\tnote{5}\\
			\midrule
			09:45\tnote{6}&9.72&9.74&9.79&9.72&$2.38\times10^{6}$&$2.32\times10^{7}$&1327&1.77&1.43&$1.45\times10^{8}$&$5.93\times10^{7}$\\
			10:00&9.61&9.73&9.73&9.60&$3.63\times10^{6}$&$3.51\times10^{7}$&1716&2.30&1.81&$1.16\times10^{8}$&$4.85\times10^{7}$\\
			\vdots&\vdots&\vdots&\vdots&\vdots&\vdots&\vdots&\vdots&\vdots&\vdots&\vdots&\vdots\\
			15:00&9.70&9.70&9.71&9.68&$2.21\times10^{6}$&$2.14\times10^{7}$&728&0.21&0.89&$8.61\times10^{7}$&$1.54\times10^{8}$\\
			\bottomrule
		\end{tabular}
		
		\begin{multicols}{2}
			\begin{tablenotes}
			\footnotesize
			\item[1]number of transactions
			\item[2]commission ratio
			\item[3]volume ratio
			\item[4]commission purchase
			\item[5]commission sale
			\item[6]Chinese stock market opening time: \\9:30$\sim$11:30 and 13:00$\sim$15:00
			\end{tablenotes}	
		\end{multicols}
	\end{threeparttable}
\end{table}	
\begin{table}[ht]
	\scriptsize
	\renewcommand\arraystretch{1.2}
	\centering
	\caption{\footnotesize Data Example 2\ \ \ \ \ \ \ \ (Stock: SH600000\ \ \ \ Date: Jan $2^{th}$, 2019)}
	\label{table3}
	\begin{tabular}{cccccccccccc}
		\toprule
		$\mathbf{Time}$&$\mathbf{Close}$&$\mathbf{Open}$&$\mathbf{Highest}$&$\mathbf{Lowest}$&$\mathbf{Volume}$&$\mathbf{Amount}$&$\mathbf{NOT}$&$\mathbf{CR}$&$\mathbf{VR}$&$\mathbf{CP}$&$\mathbf{CS}$\\
		\midrule
		11:30&9.67&9.74&9.79&9.58&$1.37\times10^{7}$&$1.32\times10^{8}$&7429&1.75&1.03&$1.13\times10^{9}$&$4.23\times10^{8}$\\
		15:00&9.70&9.70&9.71&9.68&$2.21\times10^{6}$&$2.14\times10^{7}$&728&0.21&0.89&$8.61\times10^{7}$&$1.54\times10^{8}$\\
		\bottomrule
	\end{tabular}
\end{table}
We use the first 80\% of the data set of dates as a training set and the rest 20\% as a test set to avoid data snooping. The training set and the test set are compared during the training. The accuracy of the above can be used to determine whether the model is over-fitting or not. Our data set is shared by LSTM model and CNN model in the entire section.
There are some differences in our training set: we still use $M$ as the data set. The LSTM model uses two types of data. The first is the time series of the price data for each 15 minutes of the previous day, 240 minutes or 16 steps in total, and the second is the price data for every 120 minutes in the first 10 days, which is a time series of 20 steps; CNN model uses one type of data, which is the volume data for every 15 minutes for the first five days.

\subsection{Experiment of the LSTM model}
\subsubsection{Hyperparameter Selection for Loss Function and Optimizer}
Because the model is to process classification problem, we chose the cross-entropy cost function as the loss function:
\begin{displaymath}
Loss=-\frac{1}{m}\sum_{j=1}^{m}\sum_{i=1}^{n}y_{ji}\log(\hat{y}_{ji})
\end{displaymath}

In this equation, $m$ is the total number of samples; $n$ is the total number of categories; $y_{ij}$ represents the real category (only 0 or 1), and $\hat{y}_{ij}$ is the probability of the corresponding category output by the softmax layer.

    \begin{figure}[htbp]
        \centering
        \includegraphics[width=.8\textwidth]{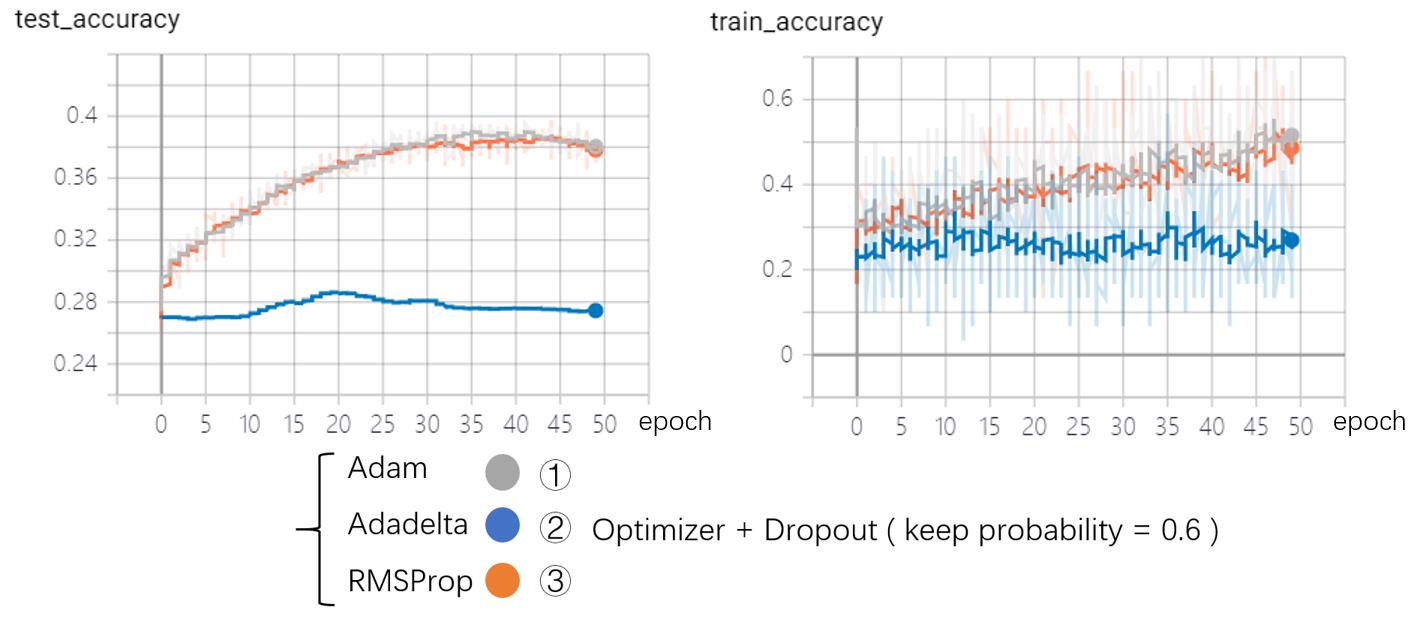}
        \caption{The Accuracy of Different Optimizer in Test Set and Train Set}
        \label{pic3}
    \end{figure}

    \begin{figure}[htbp]
        \centering
        \includegraphics[scale=0.3]{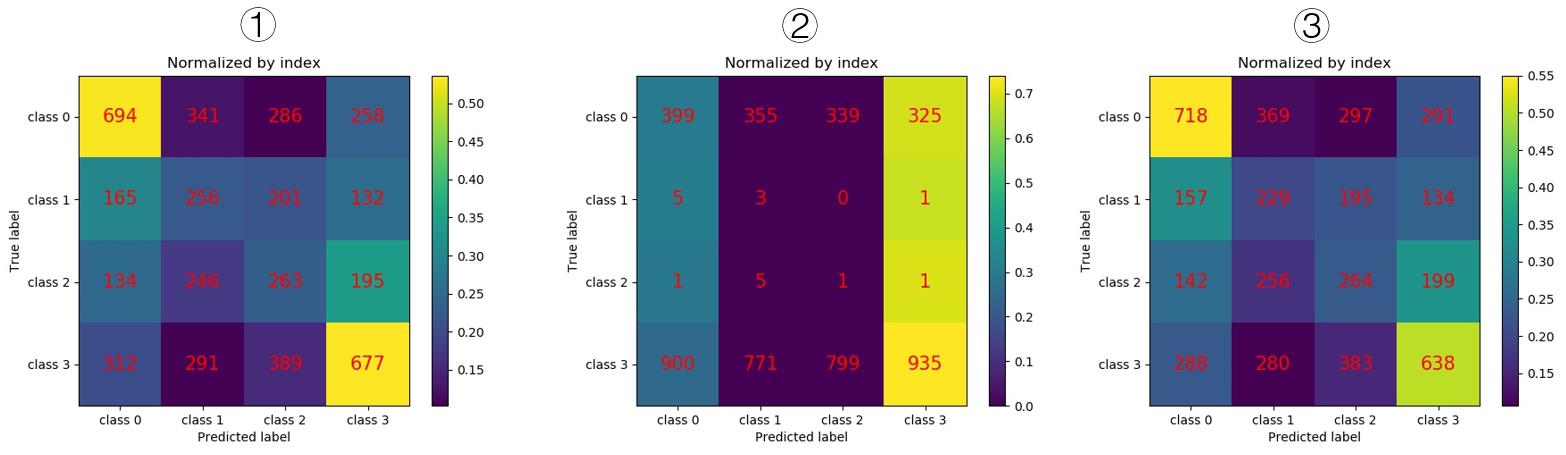}
        \caption{The Confusion Matrix Thermograms of the three models}
        \label{pic4}
    \end{figure}

For the optimizer, we selected Adam, Adadelta, and RMSProp three adaptive optimizers for testing (learning rate is 0.001). The performance of different optimizers is represented in Figure \ref{pic3}. The test uses batch gradient descent method: there are 30 samples per batch and all samples do 50 iterations. Also, all the following tests are the same.

We use the early stopping to avoid overfitting at the $30^{th}$ epoch. The Confusion matrix thermograms of the three models which are normalized by index and are represented in Figure \ref{pic4}.

As can be seen from the graph, the Adadelta optimizer is inferior, and Adam optimizer and the RMSProp optimizer are equally effective. Therefore, according to this result, we chose Adam optimizer in our model.

\subsubsection{Regularization}
   \begin{figure}[htbp]
        \centering
        \includegraphics[width=.8\textwidth]{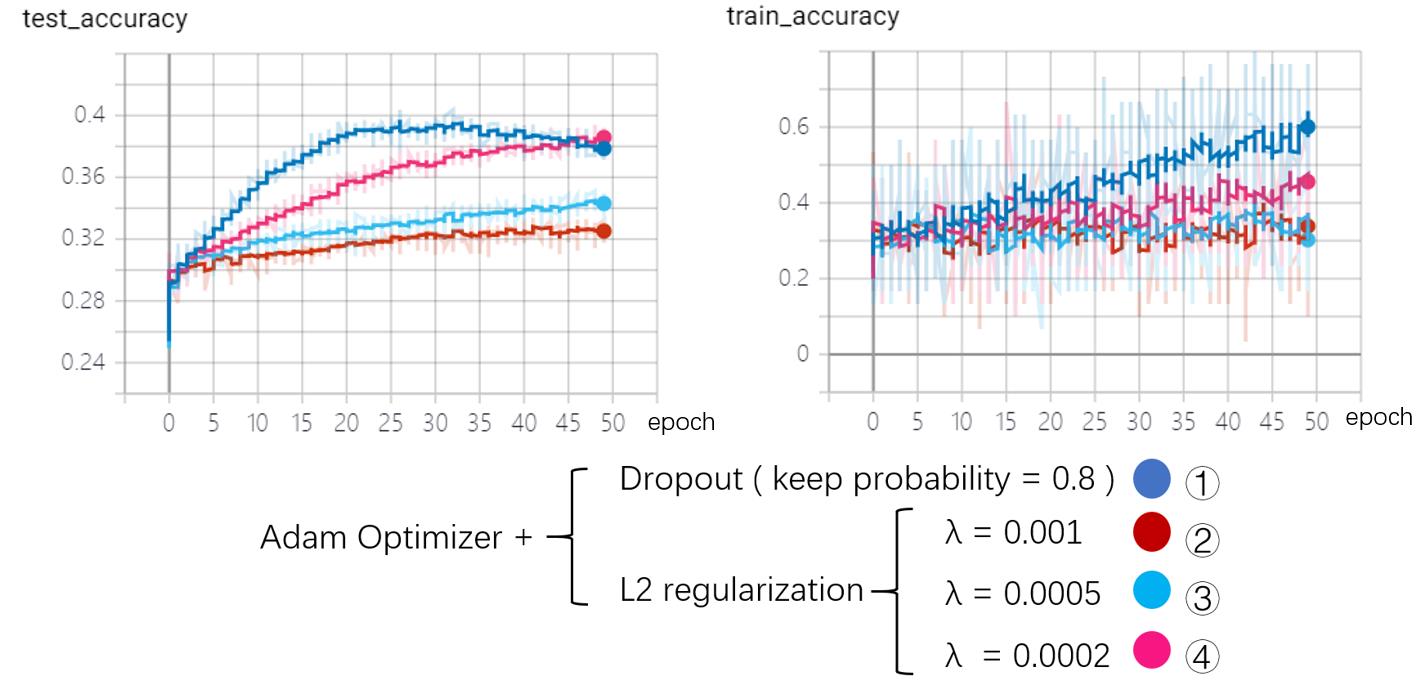}
        \caption{The Accuracy of Different Regularization in Test Set and Train Set}
        \label{pic5}
    \end{figure}
Regarding the regularization method, we compared the performance of the same data using the dropout layer and L2 regularization (weight decay). L2 regularization adds a regular term to the loss function:
\begin{displaymath}
C=C_{0}+\frac{\lambda}{2n}\sum_{w}w^{2}
\end{displaymath}

where $C$ is the objective function, $C_{0}$ is the original penalty function---the cross-entropy cost function,$w$ is the parameter of the model, $n$ is the total number of model parameters, $\lambda$ is a hyperparameter. We have tried different parameters for lambda and plot the result in Figure \ref{pic5}.

    \begin{figure}[htbp]
        \centering
        \includegraphics[width=.8\textwidth]{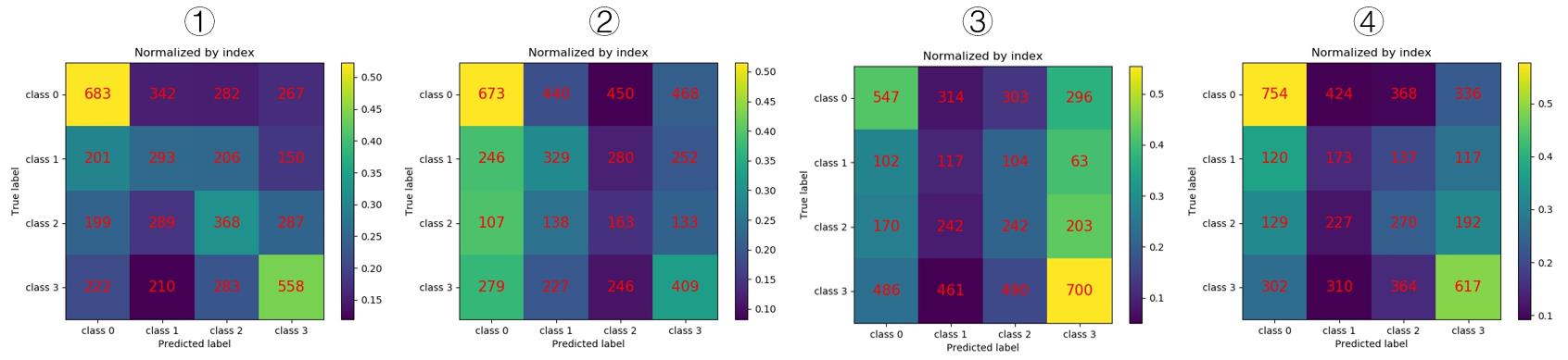}
        \caption{The Confusion Matrix Thermogram for different hyperparameter}
        \label{pic6}
    \end{figure}

Moreover, in Figure \ref{pic6}, we can see the confusion matrix thermogram for the three models at the $30^{th}$ epoch with using the early stopping to avoid overfitting is as follows, which is normalized by index.

Depend on the result showed in the graph, we can conclude that although L2 regularization can effectively avoid overfitting, using the dropout layer can make the model converge to a better solution fast. Hence, our model decides to use the dropout layer for regularization.

\subsubsection{Keep Probability of the Dropout Layer}
\begin{figure}[htbp]
        \centering
        \includegraphics[width=.8\textwidth]{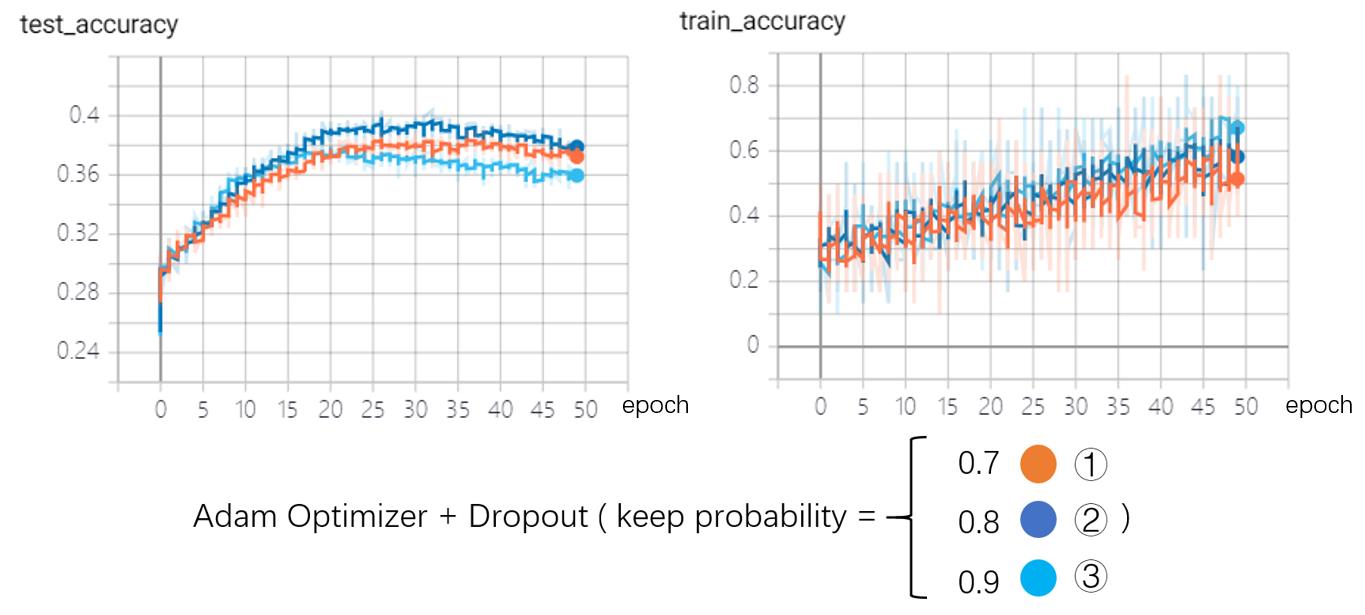}
        \caption{The Accuracy of Different Keep Parameter in Test Set and Train Set}
        \label{pic7}
    \end{figure}

Drop probability of the dropout layer
The keep probability determines the probability that each data entering the dropout layer will be retained. We use different keep parameter parameters to test on $I$ to see its effect on the model in Figure \ref{pic7}.

The confusion matrix thermogram for the three models at the $30^{th}$ epoch ,with using the early stopping to avoid overfitting, normolized by index, are represented in Figure \ref{pic8}:
    \begin{figure}[htbp]
        \centering
        \includegraphics[width=.8\textwidth]{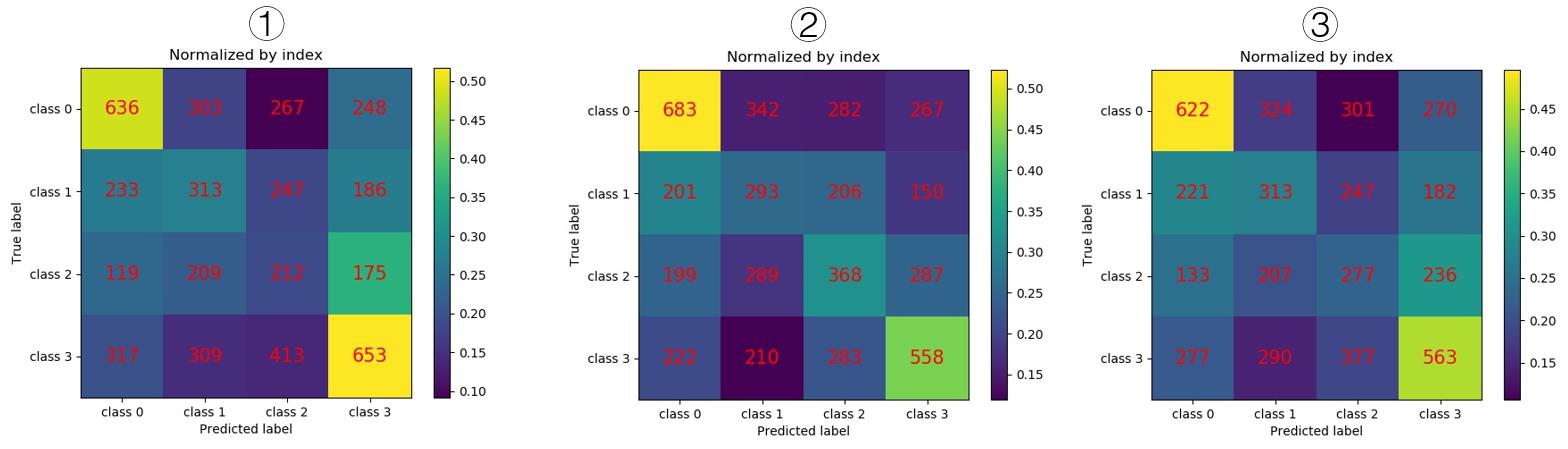}
        \caption{The Confusion Matrix Thermogram for the Three Models}
        \label{pic8}
    \end{figure}
It can be seen from the graph that when keep probability is 0.8, it is better than 0.7, 0.9. The accuracy of the model prediction is higher, so we choose keep probability$=$0.8.

\subsubsection{Training the LSTM Model}
We consider that the previous day's trading situation has a greater impact on the day, and the model should have a more sensitive and subtle understanding of it, which means that higher frequency data should be used for training. Moreover, the model should not ignore the long-term influence of fluctuations during the past several days. Then, it needs to use longer time intervals to learn the correlation between the data of the present day and several days before. Therefore, under the same network structure, we input two different forms of data, and train two prediction models, regarded as model $S$ and model $L$, that have a different emphasis on the stock market approach. And then we combine the prediction results of the two models to get the final output. The first type chooses the input data as the time series of the price data per 15 minutes of the previous day, 240 minutes in total, same as 16 steps. The training model predicts the next day's return $y$, and the second uses the price per 120 minutes in the first 10 days. The data with 20 steps predicts the next day's return $y$.

The training results are as follows:
    \begin{figure}[htbp]
        \centering
        \includegraphics[width=.9\textwidth]{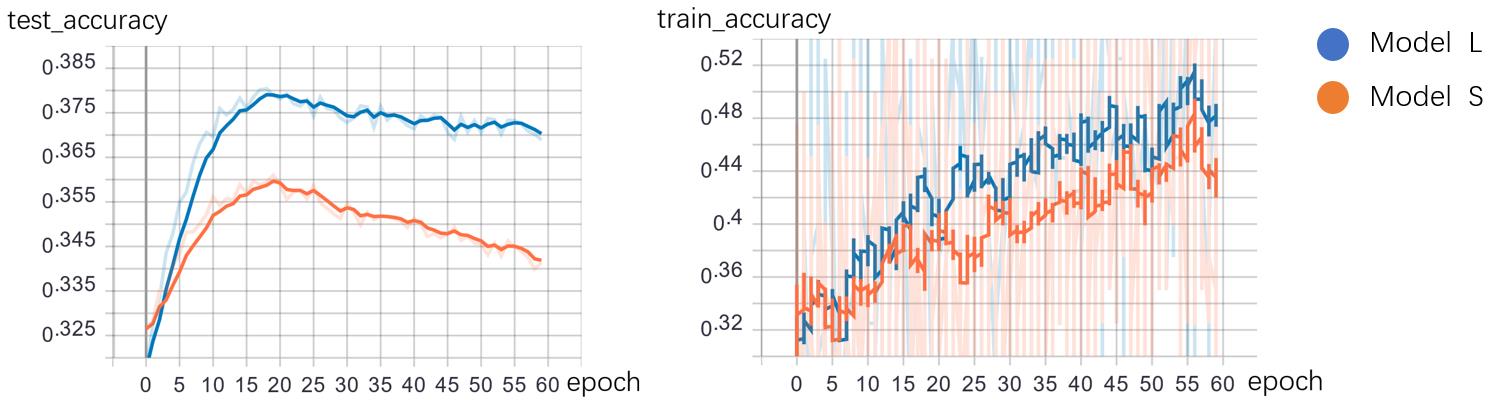}
        \caption{The Accuracy of the two models}
        \label{pic9}
    \end{figure}

\begin{figure}[htbp]
        \centering
        \includegraphics[scale=0.3]{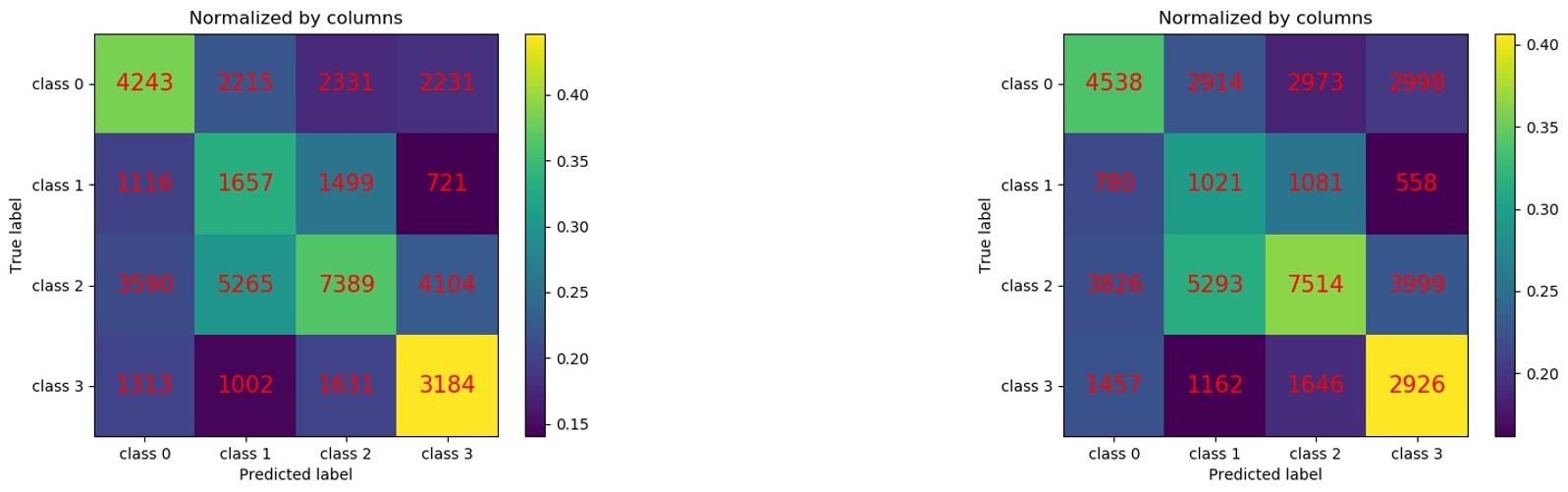}
        \caption{The Confusion Matrix Thermogram for Model S(left) and Model L(right)}
        \label{pic10}
    \end{figure}

It can be seen from the Figure \ref{pic9} that both models are valid compared to the random selection of 25\% accuracy.  And the model $L$ is better predicted than the model $S$.  The confusion matrix thermodynamics of the two models at the $20^{th}$ epoch are in Figure \ref{pic10}, which are normalized by columns.

The brighter areas of the image are concentrated near the diagonal, which again demonstrates the validity of the model. The areas in the lower right and the upper left corner are the brightest, which indicates that the model has a more accurate prediction of the situation of sharply rise and fall, specifically, the precision ratio is higher. The darker area in the middle indicates that the model does not distinguish between small rises and small falls. Overall, we use the early stop strategy to set the checkpoint to use the two models at the $20^{th}$ epoch as the final model to predict.

\subsection{Experiment of the CNN Model}
The CNN framework proposed here use the same date input as LSTM, so the data processing part is identical. Moreover, cross-entropy cost function is chosen as the loss function and Adam as the optimizer, the same as LSTM. See 3.3.1 and 3.3.2 for details. In following sections, the paper mainly focuses on the comparisons of different structures of framework, special regularization method for CNN and the expansion of input data. Experiment results are presented to help develop our theories.

\subsubsection{The Difference and Improvements from Framework by E. Hoseinzade and S. Haratizadeh}

In this section, the difference between two frameworks is explained. Experiment results are employed to conclude that the CNN framework proposed here performs better. And we try to develop theories why.
    \begin{figure}[htbp]
        \centering
        \includegraphics[width=.8\textwidth]{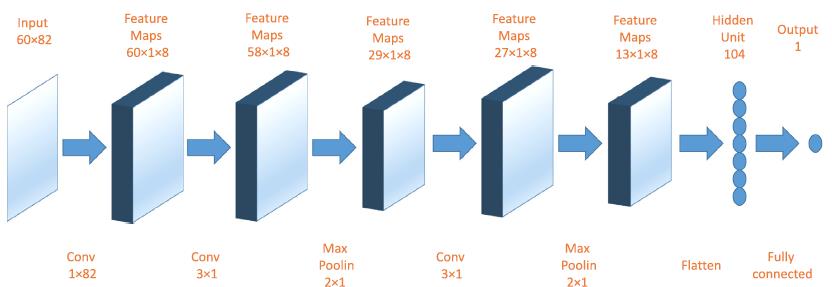}
        \caption{Ehsan Hoseinzade and Saman Haratizadeh's Framework}
        \label{ehsan}
    \end{figure}
From the framework in Figure \ref{ehsan}, it can be seen that Ehsan Hoseinzade and Saman Haratizadeh’s framework is more complicated, with 3 convolution layers, 2 pooling layers and one fully connected layer. As mentioned before, the first convolution layer has the same function as that our framework, which is to extract high-level features from primary ones. And the second convolution layer is to generate durational features by aggregating the information in consecutive time periods. So, convolution kernels of 3$\times$1 is utilized, which generate new durational features containing information from 3 consecutive time periods. Such design is inspired by the observation that most of the famous candle-stick patterns like Three Line Strike and Three Black Crows try to find meaningful patterns in three consecutive days. The third layer is a pooling layer that performs a 2$\times$1 max pooling, that is a very common setting for the pooling layers. Next, the third convolution layer is similar to the second one, intended to further extract durational features. It is followed by another same max pooling layer and a fully connected layer.
In the very beginning, we chose to apply similar framework in our task, using 2 convolution layers to extract high-level features and durational features respectively, followed by a max pooling layer and finally a fully connected layer.

    \begin{figure}[htbp]
        \centering
        \includegraphics[width=.8\textwidth]{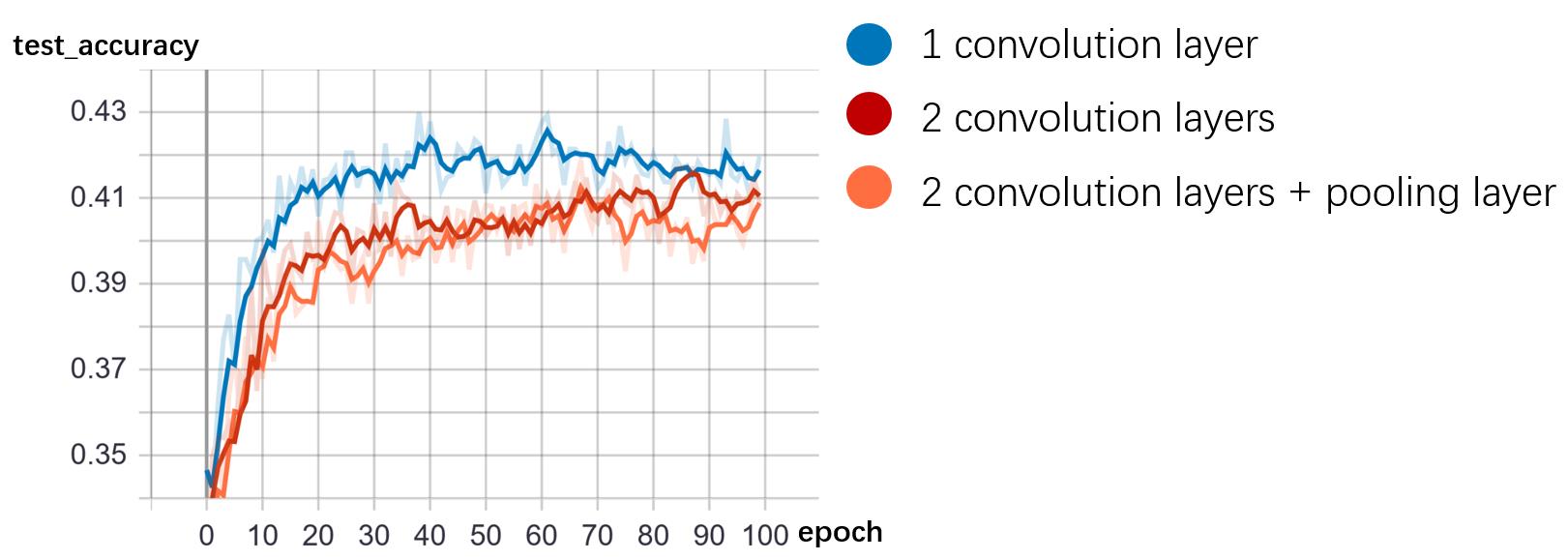}
        \caption{Accuracy of Different Framework}
        \label{1}
    \end{figure}

However, during experiments, we find that the average pooling layer performs better than a max pooling one. Better performance means higher accuracy rate in the test set in this section and those below. In Figure \ref{1}, we find that the removal of pooling layer leads to even better performance. Therefore, we assume that pooling layer is not suitable for our task, omitting too much information. So, we remove the pooling layer.
After that, we find that the removal of the second convolution layer also leads to better performance. Our assumption is as follows: The idea of such layer is inspired by candle-stick patterns like Three Line Strike. It is a good idea to combine information of past consecutive 3 days to predict daily return today. But it should be noticed here that the time span of each time period that is combined with others should match that of the time period being predicted. For instance, it may be effective to combine information of 3 consecutive 15-minute periods to predict the price direction in the next 15-minute period. But such effectiveness might not be valid when the whole daily return of today is being predicted. So, we remove the second convolution layer.
So far, there remain one convolution layer and a fully connected layer in our semi-finished framework. After experiment, we discover that adding another fully connected layer lead to better performance. Such change increases the depth of the framework and thus better capture the information hidden in the data. However, we also find that increasing the number of fully connected layers to 3 or more does not lead to obvious improvement. So, we decide to employ 2 fully connected layers, the number of neurons being 250 and 100 respectively and ReLU function as activation function.

\subsubsection{Comparison with DNN}

\begin{figure}[htbp]
        \centering
        \includegraphics[width=.8\textwidth]{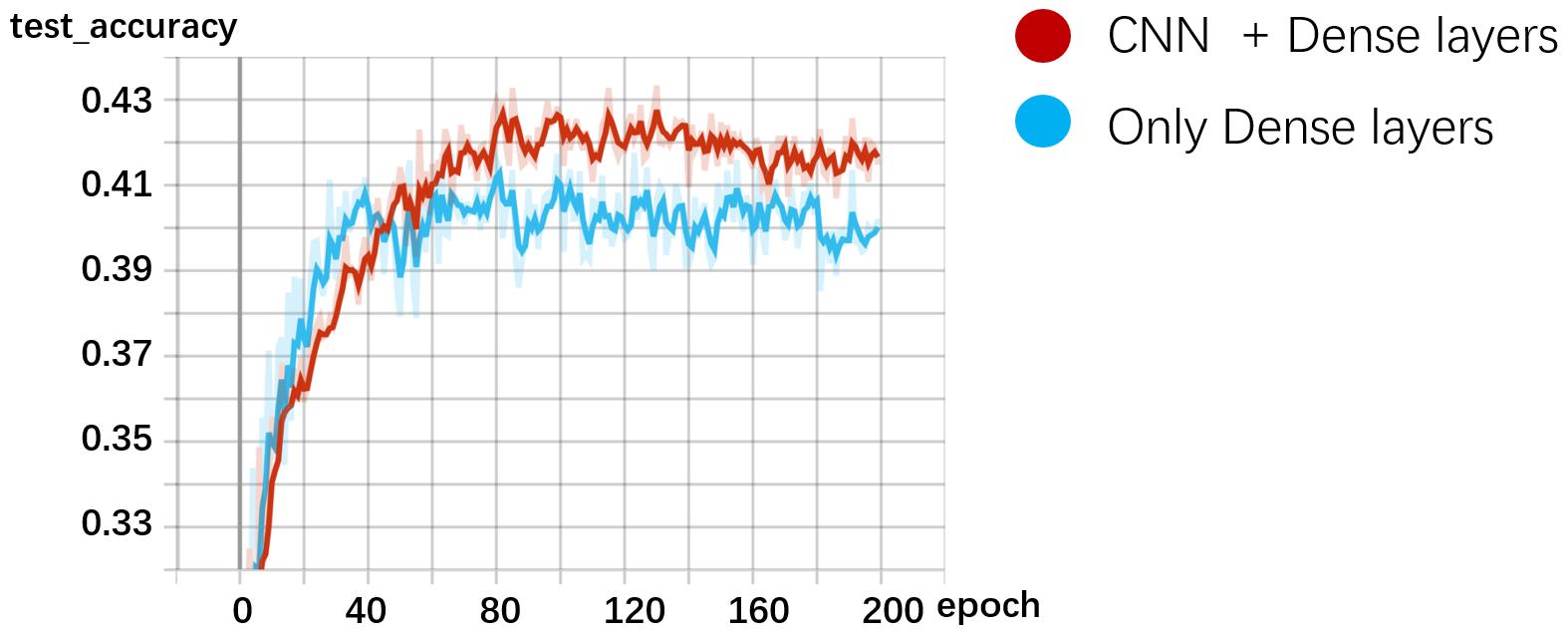}
        \caption{The Accuracy of DNN and CNN}
        \label{dnn}
    \end{figure}

We compare our convolution $+$ 2 fully connected layers framework with a framework with only 2 fully connected layers, and the result is in favor of ours. In Figure \ref{dnn}, this demonstrates that it is effective to use one convolution layer to extract high-level features.

\subsubsection{Regularization-SpatialDropout}
\begin{figure}[htbp]
        \centering
        \includegraphics[width=.5\textwidth]{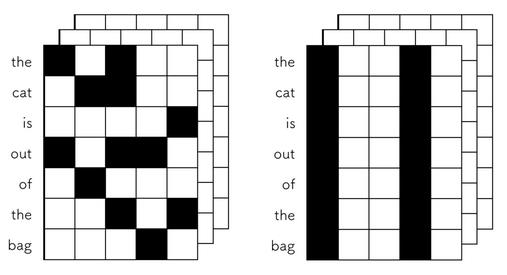}
        \caption{Ordinary Dropout vs. SpatialDropout}
        \label{ordinary}
    \end{figure}

In our CNN framework, besides commonly used dropout method in fully connected layers, our regularization methods also include SpatialDropout, a dropout method designed for CNN. Ordinary dropout method randomly chooses some components of input data without any fixed spatial patterns and turn them to zero, while SpatialDropout randomly chooses some rows or columns of input data and turn them to zero. According to Figure \ref{ordinary}, such method is proved to be effective in image identification.

In our framework, when training our model, in each batch we employ such method to randomly choose 3 columns in input data and turn them to zero. That means we drop 3 out of 11 primary price-volume features and use 8 randomly left features to construct high-level features in each batch of the training set. Through this method, the model can learn to use different combinations of 8 primary features to construct high-level ones. It enhances CNN’s ability to extract different features.
In addition, the dropout methods themselves can be comprehend as a low-cost ensemble strategy, whose essence is similar to random forest. Specific to our task, SpatialDropout forces the framework to utilize randomly left features to approach the best model. In this way, each primary feature is supposed to make contribution to the final model, which prevents our model from placing extra emphasis on some primary features in the training set, and thus, from overfitting.

\subsubsection{Input Data of Five Days VS. One Day}
    \begin{figure}[htbp]
        \centering
        \includegraphics[width=.8\textwidth]{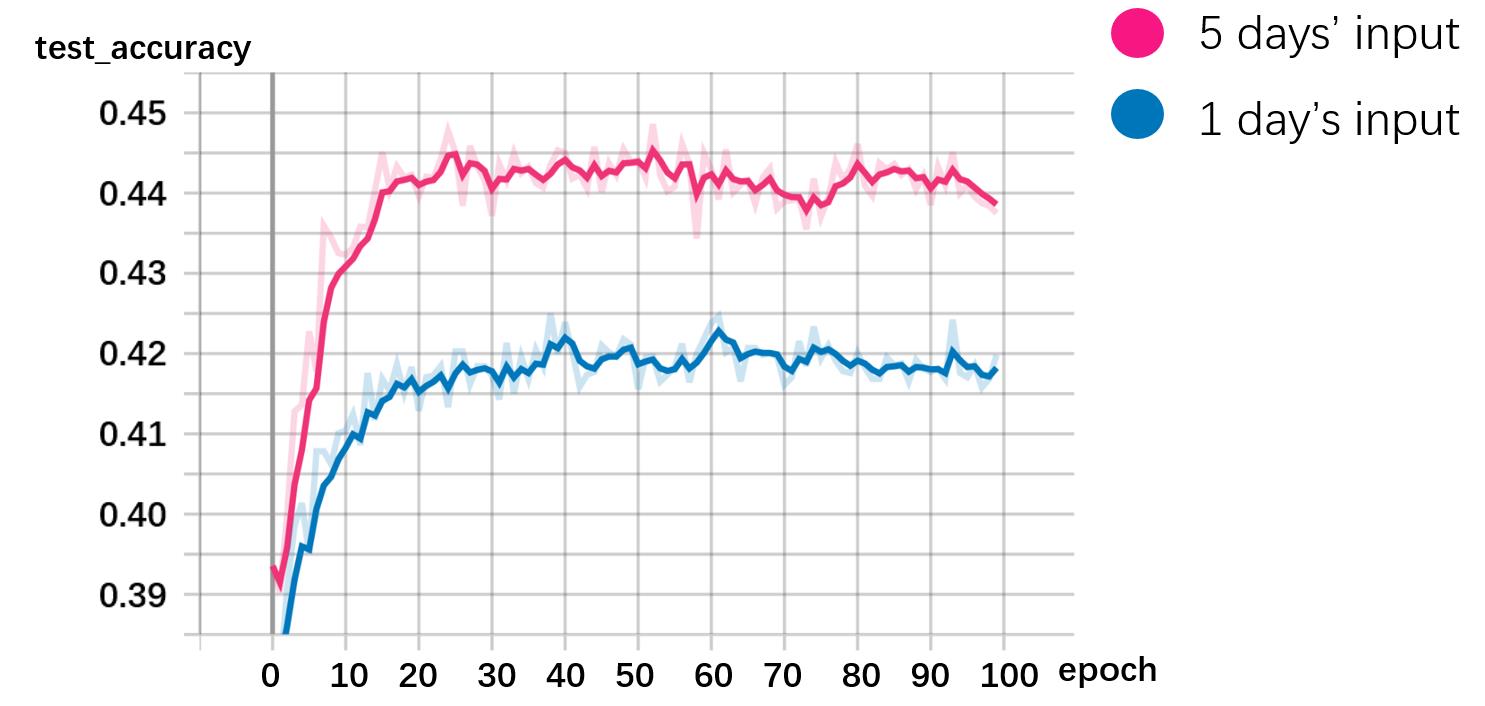}
        \caption{The Accuracy of Inputs of Different Time Span}
        \label{vs}
    \end{figure}
After accomplishing improvements mentioned above, the accuracy on the test set is still not satisfying. We assume that the input data of one day is not sufficient to make predictions on the daily return today. Therefore, we consider elongating the time span on input data. Finally, we decide to 15-minute price-volume data in 5 past consecutive days to predict the daily return today. In Figure \ref{vs}, the experiment result shows that it indeed improves the accuracy on the test set. We think this is because such elongation enables the model to better recognize the trends of features, which contributes to better predictions.

\subsubsection{Selection of Final Framework}
In the end, we determine to use CNN+2Dense as the final framework and feed it with 15-minute price-volume trading data in 5 past consecutive days.
    \begin{figure}[htbp]
        \centering
        \includegraphics[width=.6\textwidth]{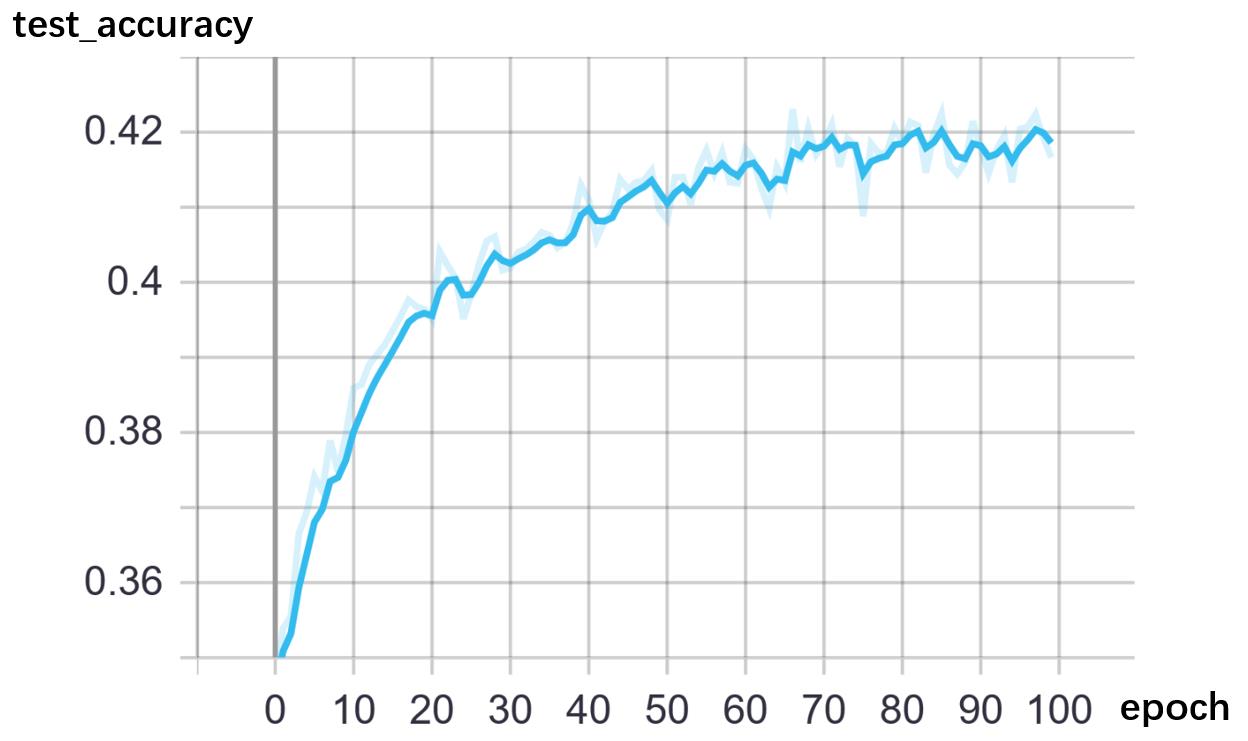}
        \caption{The Accuracy of CNN Model}
        \label{aaa}
    \end{figure}

    \begin{figure}[htbp]
        \centering
        \includegraphics[width=.6\textwidth]{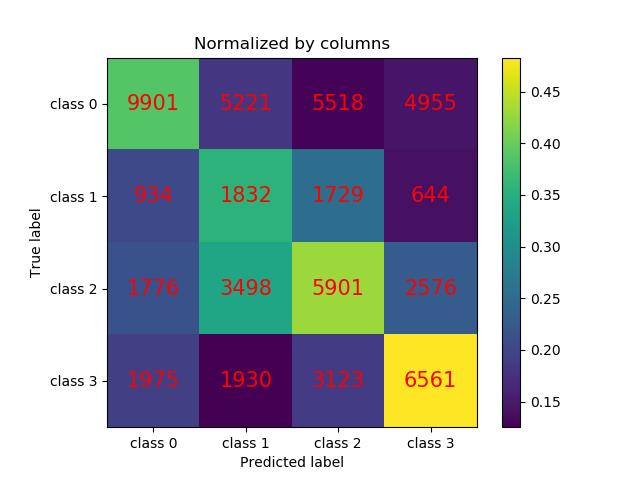}
        \caption{The Confusion Matrix Thermogram for CNN Model}
        \label{jncn}
    \end{figure}
The performance in test set is presented in (Figure \ref{aaa},\ref{jncn}) , from which we can see that the accuracy approaches 42\%, compared to 25\% if the stocks are randomly chosen.

\subsection{Make Stock Purchase Advice}
In the section of model construction, we mentioned that the softmax layer outputs the corresponding probability that the current sample belongs to four categories, and the one with the highest probability is its prediction result. In practical applications, we hope that the model not only predicts the classification of the sample but also want it can comprehensively consider the benefits and risks to directly give stock purchase advice. To achieve this purpose, we use the formula as follows:
\begin{displaymath}
E=\frac{1}{2}\sum_{model\,1,2}
p(S\in \{Class\ k\})\,\bar w(Class\ k)
\end{displaymath}
$S$ represents the input sample, and $\bar{w}(Class\ k)$ represents the mean of all the yields belonging to category $k$ in sample $J$, which we use to represent the expected rate of return for each category. $E$ represents the final prediction of the sample yield.
The formula takes into account the possible rise and fall of the sample and the predictions of the two models. After sorting the final predicted rate of return, the higher the top-ranked stock, the more recommended it is.

\section{CONCLUSION}
\subsection{Result Analysis}
\begin{figure}[b]
        \centering
        \includegraphics[width=.5\textwidth]{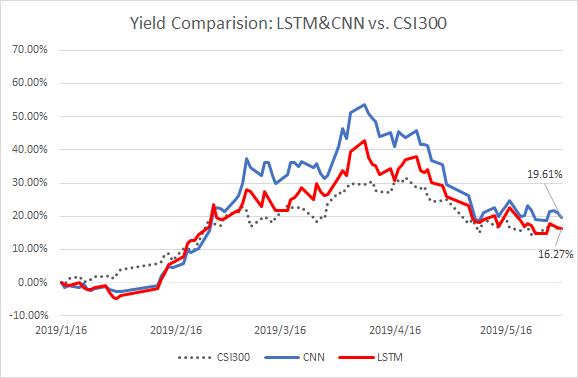}
        \caption{The Backtesting Result with Transaction Fee}
        \label{fee}
    \end{figure}

\begin{figure}[htbp]
        \centering
        \includegraphics[width=.5\textwidth]{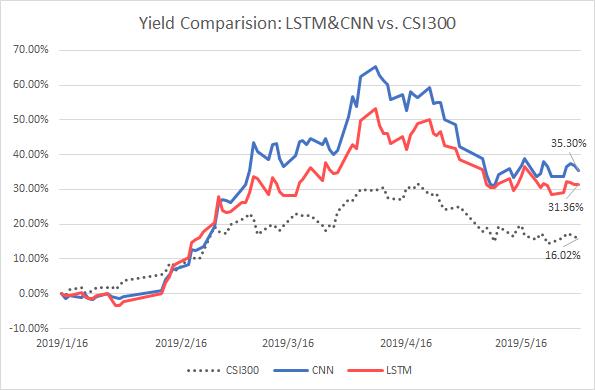}
        \caption{The Backtesting Result without Transaction Fee}
        \label{nofee}
    \end{figure}

Finally, we connect the LSTM model and the CNN model to the backtesting framework, using the CSI300 Index as the baseline, and simulate trading from January 16, 2019 to May 31,2019. The specific trading operation is as follows: regarded $I$ as the stock pool, we trade according to the purchase proposal given by the model--- considering the transaction fee, we only consider buying no more than 20 stocks with an predicted profit of 0.14\% or more. The funds are allocated on average, and the portfolio are changed daily. Our results are represented in Figure \ref{fee}, \ref{nofee} and Table \ref{table44}.

\begin{table}[h]
 \renewcommand\arraystretch{1.2}
 \centering
 \caption{Result of the backtest}
 \label{table44}
 \begin{tabular}{ccc}
  \toprule
   &$\mathbf{LSTM\ \ model}$&$\mathbf{CNN\ \ model}$\\
  \midrule
  $\mathbf{Time\ Begin}$&\multicolumn{2}{c}{2019-01-16}\\
  $\mathbf{Time\ End}$&\multicolumn{2}{c}{2019-05-31}\\
  $\mathbf{Baseline\ Simple\ Interest\ Rate}$&\multicolumn{2}{c}{$16.02\%$}\\
  $\mathbf{Transaction\ Number}$&$88$&$93$\\
  $\mathbf{Winning\ Rate}$&$51.14\%$&$51.61\%$\\
  $\mathbf{Profit\ in\ Rate}$&$31.36\%$&$35.30\%$\\
  $\mathbf{Fee\ Cost\ in\ Rate}$&$15.09\%$&$15.69\%$\\
  $\mathbf{Net\ Profit\ in\ Rate}$&$16.27\%$&$19.61\%$\\
  $\mathbf{Maximal\ Draw\ Down\ in\ Rate}$&$21.27\%$&$25.43\%$\\
  $\mathbf{Interest/MDD\ (Rate)}$&$79.48\%$&$82.91\%$\\
  \bottomrule
 \end{tabular}
\end{table}

It can be seen that both models have obvious outperform baselines regardless of the handling fee, which explains their effectiveness from another perspective. The model slightly outperforms the baseline when considering the handling fee.  And the curves of the two models are more intense than the baseline changes, which means that they tend to make more aggressive decisions when trading. Therefore, it can be concluded that our model is effective in dealing with stock return prediction with high frequency primary price-volume data.

\subsection{Future Work}
Through the above tests and analysis, we found that our two models have the advantages of fast convergence, strong generalization ability, and no need for construction factors. However, the accuracy of prediction still has a gap between our expectations. In the application of stock forecasting, LSTM and CNN are unable to achieve their superiority in the field of natural language processing and image processing. In future research, we want to utilize strengths of LSTM and CNN to construct a new model that combines CNN and LSTM. Specifically, we hope to use CNN’s ability to automatically extract high-level features and enhance important features to construct factors. After CNN generating time series about high-level factors, this time series is then used as an input to the LSTM. Moreover, LSTM’s sensitivity to time series is used to predict future stock price movements. This new model may overcome the difficulties of constructing factors and enhance our model's ability of prediction.

Also, the data used in our a set are price-volume features per 15 minutes and 120 minutes. Because of insufficient computing power, we did not use data per 1 minute or 5 minute. These higher frequency data can capture subtle information and tendency in stock market more precisely. Therefore, in the future, we can use price-volume features per 1 minute as our dataset to obtain more primary features and improve prediction accuracy.

Besides, because in our case we trade stocks everyday, it will generate a lot of unnecessary transaction fees. Considering the expense of transaction fees, we want to apply reinforce learning to maximize future profits. Moreover, it is easy to construct the environment of the complicated and sensitive stock market without constructing features by ourselves. Hence, reinforce learning is a feasible attempt in the future.

\subsection{Conclusion}
This paper applies neural network of deep learning to construct
two models of LSTM and CNN to forecast the expected return rate of the stock today, and to maximize the total return by adapting an appropriate strategy. We analyze the performances of LSTM and CNN and verify their effectiveness and rationalities for the application of forecasting stock prices. Although, these two models have overcame some difficulties, there is still a possibility of advancement such as avoiding unnecessary transaction fees. In our later works, we will focus on solving these problems.

\subsection{Acknowledgement}
We would like to express sincere appreciations to Maxwell Liu from ShingingMidas Private Fund, Xingyu Fu from Sun Yat-sen University for their generous guidance throughout the project. Also, we are grateful to Kangkang Jiang from Sun Yat-sen University for his assistance all the way. Without their supports, we cannot complish such a challenging task.

\newpage
\nocite{*}
\bibliographystyle{plain}
\bibliography{main}
\end{document}